\PassOptionsToPackage{dvipsnames}{xcolor}
\documentclass[aip,apl,reprint,a4paper,superscriptaddress,floatfix,amsmath,amssymb,amsfonts,noshowpacs,longbibliography]{revtex4-2}

\usepackage{newtxtext,newtxmath}
\usepackage[T1]{fontenc}
\usepackage[utf8]{inputenx}
\input{ix-utf8enc.dfu}
\usepackage{graphicx}
\usepackage{sidecap}
\sidecaptionvpos{figure}{t}
\usepackage[dvipsnames]{xcolor}
\usepackage{soul} 
\usepackage{xspace}
\usepackage{siunitx}
\usepackage{url}
\usepackage{floatrow}

\usepackage[
,textwidth=17.5cm
,textheight=23.7cm
,verbose
,pdftex
]{geometry}
\usepackage{xspace}

\usepackage[pdftex]{hyperref}
\hypersetup{
	pdftitle={Optical properties of ScN layers grown on Al$_2$O$_3$(0001) by plasma-assisted molecular beam epitaxy},
	colorlinks,
	citecolor=blue,
	linkcolor=blue,
	urlcolor=blue
}

\begin{document}

\title{Optical properties of ScN layers grown on Al$_{\boldsymbol{\mathsf{2}}}$O$_{\boldsymbol{\mathsf{3}}}$(0001) by plasma-assisted molecular beam epitaxy}

\author{Duc V. Dinh}
\email[Electronic email: ]{dinh@pdi-berlin.de}
\affiliation{Paul-Drude-Institut für Festkörperelektronik, Leibniz-Institut im Forschungsverbund Berlin e.V., Hausvogteiplatz 5--7, 10117 Berlin, Germany.}

\author{Frank Peiris}
\affiliation{Paul-Drude-Institut für Festkörperelektronik, Leibniz-Institut im Forschungsverbund Berlin e.V., Hausvogteiplatz 5--7, 10117 Berlin, Germany.}
\affiliation{Department of Physics, Kenyon College, Gambier, 43022 Ohio, United States}

\author{Jonas Lähnemann}

\author{Oliver Brandt}
\affiliation{Paul-Drude-Institut für Festkörperelektronik, Leibniz-Institut im Forschungsverbund Berlin e.V., Hausvogteiplatz 5--7, 10117 Berlin, Germany.}


\begin{abstract}
An accurate knowledge of the optical constants (refractive index $n$ and extinction coefficient $k$) of ScN is crucial for understanding the optical properties of this binary nitride semiconductor as well as for its use in optoelectronic applications. Using spectroscopic ellipsometry in a spectral range from far infrared to far ultraviolet (0.045--8.5\,eV), we determine $n$ and $k$ of ScN layers grown on Al$_2$O$_3$(0001) substrates by plasma-assisted molecular beam epitaxy. Fits of ellipsometry data return the energies of four oscillators representing critical points in the band structure of ScN, namely, 2.03, 3.89, 5.33, and 6.95\,eV. As the infrared range is dominated by free carriers, the vibrational properties of the layers are examined by Raman spectroscopy. Despite the rocksalt structure of ScN, several first-order phonon modes are observed, suggesting a high density of point defects consistent with the high electron density deduced from Hall measurements. Finally, photoluminescence measurements reveal an emission band slightly above the lowest direct bandgap. We attribute the redshift of the peak emission energy from 2.3 to 2.2\,eV with increasing layer thickness to a reduction of the O concentration in the layers.
\end{abstract}

                  
\maketitle


The rocksalt semiconductor ScN has been first synthesized more than five decades ago.\cite{Sclar1964May,Dismukes1970Dec,Dismukes1972May} More recently, ScN has attracted much interest in conjunction with the conventional group-III nitrides, i.e., GaN and AlN. In particular, the ternary alloy (Sc,Al)N holds great potential for applications in surface acoustic wave devices,\cite{Akiyama2009Feb,Hashimoto2013Mar} field-effect transistors,\cite{Hardy2017Apr,Wang2021Aug,Dinh2023Apr} and as novel ferroelectric material.\cite{Wang2022Jul,Wang2023Jan} Recently, high-quality ScN has been grown using plasma-assisted molecular beam epitaxy (PAMBE)\cite{Lupina2015Nov, Casamento2019Oct} and hydride vapor phase epitaxy,\cite{Oshima2014Apr} enabling new applications of pure ScN. In fact, ScN has been theoretically and experimentally proposed for electronic,\cite{Adamski2019Dec} thermoelectric\cite{Rao2020Apr} and infrared optoelectronic applications.\cite{Maurya2022Jul}

In terms of its electronic properties, both experimental and theoretical studies have shown that ScN has an indirect gap at about 0.8--0.9\,eV,\cite{Saha2010Feb,Deng2015Jan,Mu2021Aug} a direct gap at 1.91--3.1\,eV at the $X$ point,\cite{Dismukes1972May,Gall1998Jul,Gall1998Dec,Gall2001Mar,Smith2001Aug, Deng2015Jan,Bai2001May,Al-Brithen2004Jul,Moram2008Oct,Saha2010Feb,Saha2013Aug,Oshima2014Apr,Lupina2015Nov,Rao2020Apr} and a direct gap of 3.58--3.75\,eV at the $\Gammaup$ point in the band structure of ScN.\cite{Deng2015Jan,Al-Brithen2004Jul,Mu2021Aug} Room-temperature photoluminescence measurements of ScN show an emission band at about 2.2--2.3\,eV corresponding to the lowest direct gap.\cite{Saha2013Aug,Lupina2015Nov,Saha2017Jun} Compared to GaN and AlN, the optical constants (refractive index $n$ and extinction coefficient $k$) of ScN are less well known, with $n$ having been most commonly extracted from transmittance\cite{Dismukes1972May,Gall2001Mar, Smith2001Aug,Bai2001May,Al-Brithen2004Jul} and reflectance measurements.\cite{Gall2001Mar,Smith2001Aug,Al-Brithen2004Jul,Deng2015Jan} While there is a good agreement between $n$ extracted from reflectance data for photon energies of 1--2\,eV and density-functional-theory (DFT),\cite{Deng2015Jan} it is important to extend these measurements to the spectral range where ScN starts to absorb strongly, i.e., above its lowest direct band gap.

Spectroscopic ellipsometry (SE) is a technique ideally suited for this task because it allows us to derive the complex refractive index or dielectric function from a single measurement, without the need of an intensity reference, or the Kramers–Kronig transformation required for transmittance or reflectance data alone. SE is the method of choice for an accurate determination of both $n$ and $k$ over a wide spectral range, and is thus frequently used to extract reliable information on phononic and electronic transitions that can be compared with the predictions of DFT. In this way, SE provides important insights into the phonon and electron dispersion relationships of the material under investigation.\cite{Lautenschlager1987Jun,Roeseler1990} Consequently, SE was employed by several groups to determine the dielectric function of ScN, but only in a limited spectral range.\cite{Gall2001Mar,Jarrendahl1998Feb,Maurya2022Jul,Saha2013Aug} Most importantly, a critical comparison of the dielectric function acquired over a wide spectral range with the predictions of modern DFT calculations has yet to be performed.

In this letter, we determine the dielectric function of ScN layers grown on Al$_2$O$_3$(0001) substrates by PAMBE using SE over a spectral range of 0.045--8.5\,eV. The data are fit with an parametric model to accurately determine the energies of the critical points in the band structure of ScN, and compared to the results of recent DFT calculations utilizing improved exchange correlation functionals. Additional information on the phononic properties is obtained by Raman spectroscopy. Finally, photoluminescence spectroscopy is used to shed light on the origin of the high electron density in the layers. 

\begin{figure*}
	\includegraphics[width=\columnwidth]{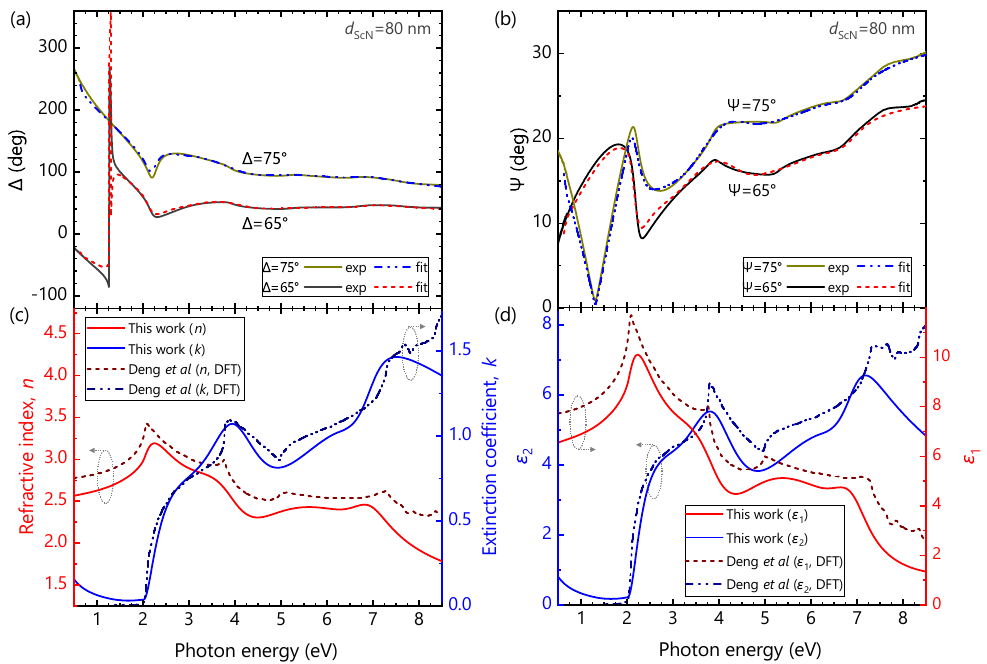}
	\caption{(a) $\Delta$ and (b) $\Psi$ spectra of the 80-nm-thick ScN layer and their point-to-point fits. (c) Real ($n$) and imaginary ($k$) parts of the complex refractive index as obtained from the fits. (d) Corresponding real ($\epsilon_1$) and imaginary ($\epsilon_2$) parts of the complex dielectric function. For comparison, (c) and (d) also show the theoretical curves (DFT) obtained in Ref.~\citenum{Deng2015Jan}.}
	\label{fig:Ellip}
\end{figure*}

ScN layers are grown by PAMBE on as-received Al$_2$O$_3$(0001) substrates (either on quarters or full 2-inch wafers). Prior to ScN growth, the substrates were outgassed for two hours at 500\,°C in a load-lock chamber attached to the MBE system. The MBE growth chamber is equipped with a high-temperature effusion cell to provide Sc metal (99.999\,\% pure Sc). A Veeco UNI-Bulb radio-frequency plasma source is used for the supply of active nitrogen (N$^*$). The N$^*$ flux is calculated from the thickness of a GaN layer grown under Ga-rich conditions, and thus with a growth rate limited by the N$^*$ flux. The Sc flux is obtained from the thickness of thin ScN calibration layers ($d\,<\,$40\,nm) measured by x-ray reflectivity. The growth temperature as measured by a thermocouple is set to 700\,°C. The ScN layers are grown with thicknesses ranging from 8 to 250\,nm under N$^*$-rich conditions. Information about the structural properties, lattice constants and surface morphology of the layers is provided in Figs.\,S1--S4 in the Supplementary Material. 

SE measurements are carried out at room temperature for two incident angles of 65 and 75° (with respect to the normal of the sample) using two Woollam ellipsometers, which have a combined spectral range from the far infrared (IR) to the far ultraviolet, specifically, 0.045--8.5\,eV. The optical constants of the ScN layers are determined by fitting the ellipsometry data with a parametric model that represents the electronic and phononic contributions as a collection of oscillators. Raman and photoluminescence (PL) spectra of the layers are recorded at room temperature using a Horiba LabRAM HR Evolution. The samples are excited with a 405-nm laser diode focused onto their surface with a spot diameter of about 2\,$\muup$m (power density $\approx$25\,kW/cm$^2$).

In ellipsometry, the amplitude ratio $\Psi$ and the phase shift $\Delta$ between light polarized parallel and perpendicular to the plane of incidence are measured.\cite{Azzam1977} Because these measured ellipsometry parameters depend on the thickness and optical constants of each layer of a structure, we develop a model that varies the thicknesses and optical constants of each layer, and change the model parameters to obtain a best fit for the experimental spectra. While the optical constants in the transparent range are well represented by a Cauchy relationship, they are modeled in the opaque range by a collection of oscillators, each representing a resonance associated with an electronic transition.\cite{Fujiwara2007}

Figures~\ref{fig:Ellip}(a) and \ref{fig:Ellip}(b) show $\Delta$ and $\Psi$ spectra for the entire spectral range measured for the 80-nm-thick ScN layer, respectively. Spectra in the IR range between 0.045 and 0.15\,eV are shown in Fig.\,S5 of the Supplementary Material. Furthermore, Fig.\,S6 of the Supplementary Material presents a comparison with a 250-nm-thick ScN layer and experimental data taken from Ref.~\citenum{Maurya2022Jul}. To fit the experimental spectra, we develop a three-layer model consisting of the Al$_2$O$_3$ substrate, the ScN layer and a top layer accounting for surface roughness modeled by a Bruggemann effective medium approximation.\cite{Aspnes1990chapter} The dielectric function of Al$_2$O$_3$ was obtained by measuring a pristine Al$_2$O$_3$(0001) substrate, and is found to compare well with results reported in the literature.\cite{Malitson1962Dec,Yao1999May} In order to determine the thicknesses of the ScN and the roughness layers, we use a narrow spectral range between 1 and 2\,eV of the measured spectra. Because ScN is essentially transparent in this range, its optical constants are represented by a Cauchy relationship (with a near-zero $k$-value).\cite{Fujiwara2007} As the number of fitting parameters is low, this method allows us to obtain unique values for the thicknesses of the ScN (81.5\,nm) and the roughness layer (8.8\,nm).

Following the determination of the thicknesses of the layers and $n$ in the transparent range, we perform a simultaneous wavelength-by-wavelength fit of $\Delta$ and $\Psi$ for the entire spectral range. This procedure may, in general, produce $n$ and $k$ values that do not obey the Kramers-Kronig relation. In the present work, the $n$ and $k$ values obtained from the wavelength-by-wavelength fit are used to construct Kramers-Kronig consistent-oscillators that represent the actual optical constants of ScN.\cite{Fujiwara2007} To fit the ellipsometry data, we use a five-oscillator model (see Fig.\,S7 in the Supplemental Material), where four oscillators represent band-to-band electronic transitions (Tauc-Lorentz oscillators\cite{Jellison1996Jul}) associated with the critical points (i.e., high-symmetry points) of the band structure of ScN and the fifth represents free electron absorption (i.e., a Drude oscillator\cite{Fujiwara2007}). A slow variation of the baseline is accounted for by two additional oscillators below and above the experimental spectral range.

Figure \ref{fig:Ellip} shows the results of the fits in terms of (c) the complex refractive index $\tilde{n} = n + ik$ and (d) the complex dielectric function $\epsilon = \tilde{n}^2 = \epsilon_1 + i\epsilon_2$. Each panel also depicts the corresponding theoretical predictions obtained from density functional calculations using the Heyd-Scuseria-Ernzerhof (HSE06) hybrid exchange correlation functional.\cite{Deng2015Jan}  As commonly observed, the experimental features are broadened compared to the theoretical ones, reflecting a certain degree of disorder in the material. In the present case, the high density of point defects in the ScN layer is the primary source for this disorder as discussed in more detail below. However, the overall agreement of experiment and theory is very satisfactory, except for a notable deviation at energies higher than 7.5\,eV. At these energies, the penetration depth of light is reduced to about 10\,nm, comparable to the peak-to-valley roughness of the layer under investigation. Hence, the optical constants measured in this spectral range are those of an effective medium rather than being representative for bulk ScN.   

The lowest oscillator at 2.03\,eV corresponds to the lowest energy direct transition at the $X$ point. The other three oscillators are associated with direct transitions that occur at the $\Gammaup$ point between the degenerate heavy- and light-hole bands and the first, second and third conduction bands, and have energies of 3.89, 5.33, and 6.95\,eV, respectively. Interestingly, the oscillator strength of the transition with an energy of 5.33\,eV is considerably higher than that of the other two transitions (see Fig.\,S7 in the Supplemental Material), indicating that it may encompass more than a single transition. Upon closer scrutiny of the band structure of ScN,\cite{Deng2015Jan} we notice that the degeneracy between the heavy- and light-hole bands is lifted at the $X$ point. The energy difference between the light-hole band and the conduction band at the $X$ point is basically equal to the one between the valence band and the second conduction band at the $\Gammaup$ point. The oscillator at 5.33\,eV thus represents a superposition of two direct transitions. 

\begin{figure}[t]
	\includegraphics[width=0.9\columnwidth]{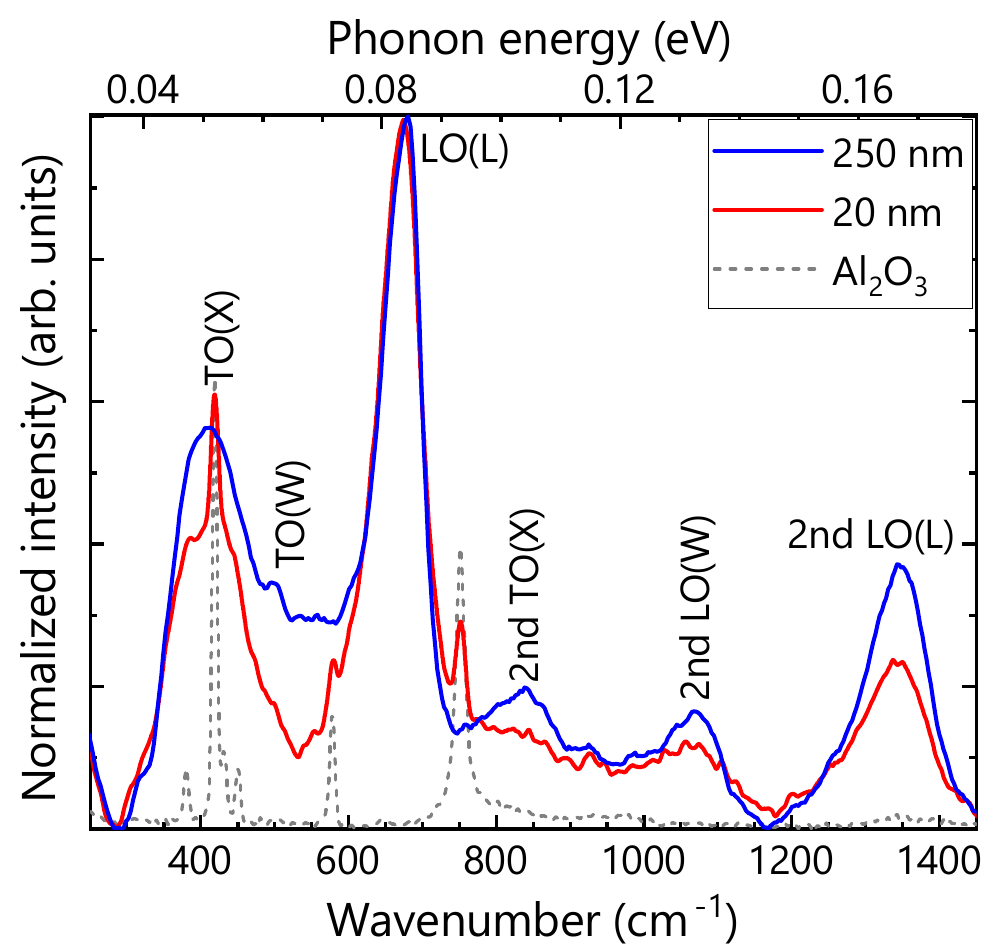}
	\caption{Raman spectra of 20- and 250-nm-thick ScN layers on Al$_2$O$_3$(0001) acquired at room temperature. The dotted line shows the Raman spectrum of the Al$_2$O$_3$ substrate for comparison.}
	\label{fig:Raman}
\end{figure}

The values of $k$ or $\epsilon_2$ in the spectral range below 2\,eV deviate significantly from the theoretical values [Figs.~\ref{fig:Ellip}(c)--\ref{fig:Ellip}(d)], which are obtained assuming an intrinsic semiconductor devoid of any doping.\cite{Deng2015Jan} In reality, ScN layers are invariably degenerately $n$-type doped, inducing a significant free-carrier absorption in the IR spectral range. This contribution to the dielectric function is taken into account in our model by a Drude oscillator, which governs its behavior in the transparent range below $\approx$2\,eV. The best fit of the SE data, assuming an effective electron mass of ($0.35\pm0.05$)$m_{0}$, return an electron density $n_e$ of 1.1$\times$10$^{20}$\,cm$^{-3}$ and a mobility of 3.4\,cm$^2$V$^{-1}$s$^{-1}$, consistent with Hall-effect measurements on this layer (performed in the van der Pauw configuration at room temperature). 

Previous reports have indicated that the transverse optical (TO)\cite{Travaglini1986Sep,Paudel2009Feb,Maurya2022Jul} and the longitudinal optical (LO)\cite{Maurya2022Jul} phonon modes in ScN are located at 360--365\,cm$^{-1}$ and 685\,cm$^{-1}$, respectively, or 0.045 and 0.085\,eV, which is within the range we have measured in the IR range (see Fig.\,S5 in the Supplemental Material).  However, the strong free-carrier absorption, as well as the strong phonon resonances of the Al$_2$O$_3$ substrate,\cite{Schubert2000Mar} dominate the optical response for very low photon energies. In fact, the Drude oscillator is perfectly adequate for fitting the experimental data as seen in Fig.\,S7 in the Supplemental Material, with no need to include additional oscillators to account for the phonon modes of ScN.      

We thus investigated the vibrational properties of the ScN layers by Raman spectroscopy. Generally, for materials with rocksalt structure, first-order Raman scattering by optical phonon modes is forbidden. Point defects such as vacancies and substitutional impurities may break this symmetry, but also structural defects such as twin boundaries.\cite{Xinh1965Nov,Travaglini1986Sep,Todorov2011Jun} In the Raman spectrum shown in Fig.~\ref{fig:Raman}, some phonon modes from Al$_2$O$_3$ (e.g., at 418 and 750\,cm$^{-1}$) are detected for the 20-nm-thick layer, but are no longer present once the layer thickness exceeds the penetration depth of the exciting laser. Hence, the remaining phonon modes can be assigned to ScN. Several first-order modes at the $X$, $W$ and $L$ points of the phonon band structure of ScN are observed, which we assign to the TO(X) at 420\,cm$^{-1}$, TO(W) at 500\,cm$^{-1}$, LO(W) at 535\,cm$^{-1}$ and LO(L) mode at ($678\pm2$)\,cm$^{-1}$, consistent with experiments conducted on bulk ScN crystals,\cite{Travaglini1986Sep} untwinned ScN layers,\cite{Lupina2015Nov} as well as with theory.\cite{Gall2001Oct,Paudel2009Feb} The prominence of zone-boundary modes, particularly the LO(L), has been discussed in detail by Paudel and Lambrecht.\cite{Paudel2009Feb}

The LO(L) spectral position is independent of the layer thickness, indicating that all of our layers are essentially fully relaxed as also suggested by x-ray diffractometry revealing a very small residual strain (see Fig.\,S3 in the Supplemental Material). This virtually full plastic relaxation is expected because the lattice mismatch between ScN(111) and Al$_2$O$_3$(0001) is huge [16\% according to the relaxed lattice constants of ScN\cite{Gall1999Nov,Moram2006Jul,Travaglini1986Sep} and Al$_2$O$_3$(0001)\cite{Campbell1961}].

\begin{figure}[t]
	\includegraphics[width=\columnwidth]{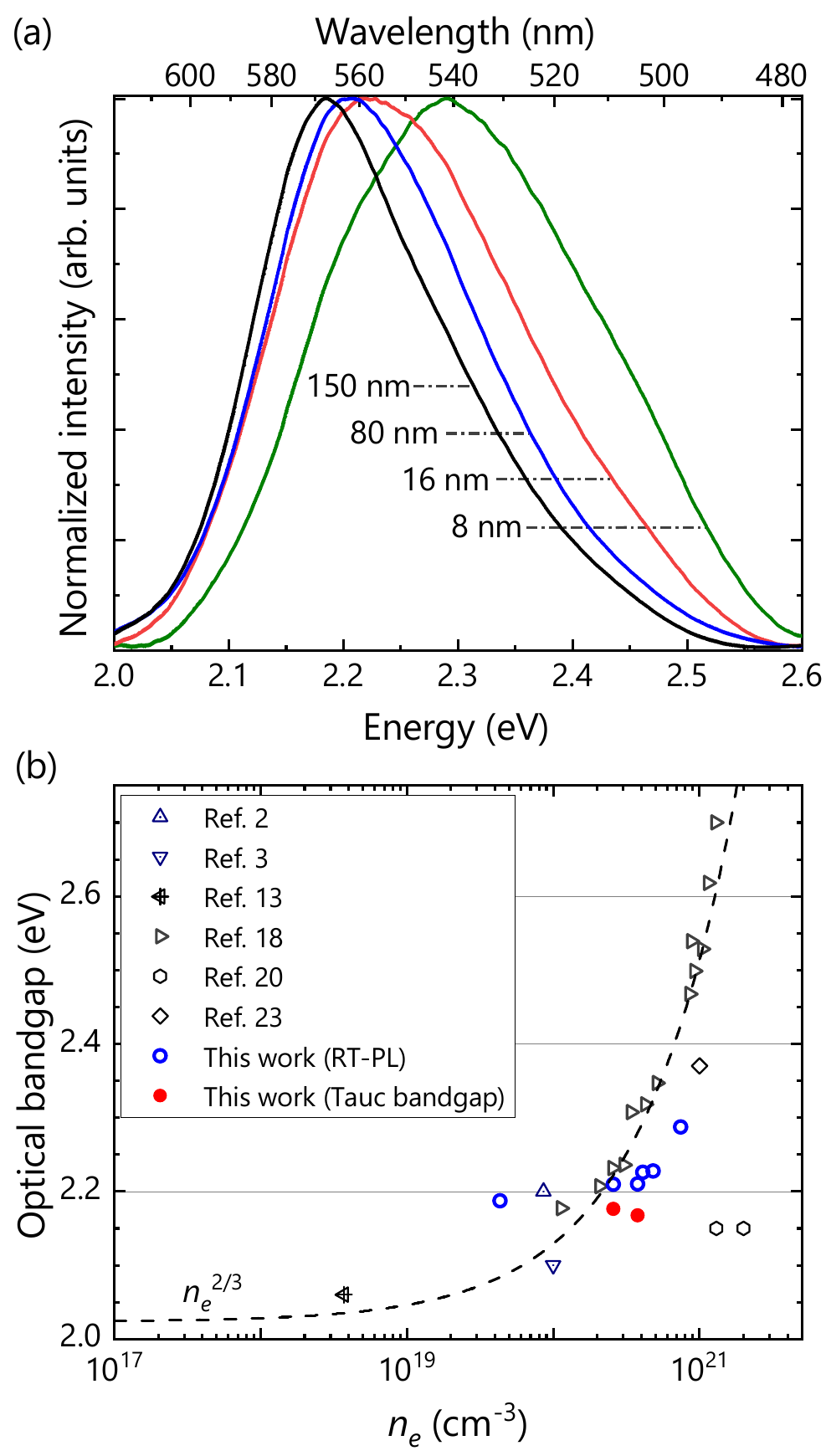}
	\caption{(a) PL spectra of ScN layers with different thicknesses acquired at room temperature. (b) PL peak energy and optical bandgap of ScN as a function of background electron concentration. The dashed line shows an $n_e^{2/3}$ dependence as expected for a free carrier gas.} 
	\label{fig:PL}
\end{figure}

Figure~\ref{fig:PL}(a) shows the PL spectra of the ScN layers under investigation. Analogously to previous reports in the literature,\cite{Saha2013Aug,Lupina2015Nov} the PL spectra of our ScN layers exhibit a broad band close to the lowest direct band-gap at the $X$ point. Direct-gap PL of indirect semiconductors has been observed before, and was found to be much enhanced for heavily doped layers.\cite{Wagner_1984,ElKurdi2009May} This emission band stems, in general, from the recombination of majority carriers with nonthermalized minority carriers at the same point of the Brillouin zone. For the particular case of ScN, the commonly reported electron densities in excess of 10$^{20}$\,cm$^{-3}$ result in the Fermi level entering the conduction band and a corresponding band filling.\cite{Deng2015Jan,Kumagai_2018} The degenerate doping thus facilitates the recombination of electrons populating the conduction-band minimum with hot photogenerated holes in the uppermost valence-band at the $X$ point.

The emission energy of the spectra depicted in Fig.~\ref{fig:PL}(a) is seen to monotonically decrease from 2.3 to 2.2\,eV with increasing layer thickness, very similar to the observation reported in Ref.~\citenum{Lupina2015Nov}. Furthermore, the width of the band decreases notably (see Fig.\,S8 in the Supplemental Material for details). While the layers experience a reduction in compressive strain from 0.3 to 0.1\% (see Fig.\,S3 of the Supplementary Material), the effect of the decreasing strain on the transition energy is much smaller than that observed. At the same time, Hall-effect measurements reveal a decrease in electron density $n_e$ from 7.4$\times$10$^{20}$\,cm$^{-3}$ to 4.3$\times$10$^{19}$\,cm$^{-3}$ with increasing layer thickness with a corresponding reduction in band filling.

Figure~\ref{fig:PL}(b) shows the dependence of the optical band gap of ScN on $n_e$ as reported in the literature and observed by us using spectroscopic ellipsometry. The optical band gap is obtained by Tauc plots\cite{Gall1998Jul} with the absorption coefficients either derived from transmittance and reflectance measurements or, in our case, directly obtained from the extinction coefficient $k$. The strong blueshift observed for $n_e > 10^{20}$\,cm$^{-3}$ is due to band filling (Burstein-Moss effect) and follows the theoretically expected $n_e^{2/3}$ dependence.\cite{Deng2015Jan,Kumagai_2018} While this shift would not be observable in emission for an ideal crystal, the presence of a high density of ionized donors induces disorder that relaxes $\mathbf{k}$ conservation, allowing the recombination of electrons and holes occupying higher energy states with different $\mathbf{k}$ vectors.\cite{Olego1980Jul,Valcheva2006Jan} For comparison, Fig.~\ref{fig:PL}(b) also shows the peak energy of the emission bands depicted in Fig.~\ref{fig:PL}(a). Note, however, that this comparison is not straightforward: while absorption sets in at the Fermi energy, the PL lineshape depends on occupation, and is here complicated by the very large high-energy broadening originating from the highly nonthermal hole distribution.\cite{Wagner_1984} The peak energy thus does not, in general, correspond to the Fermi energy.\cite{Olego1980Jul} Still, the PL peak energies are in overall agreement with the optical gap, and provide a very convenient and direct means to obtain information about the degeneracy of ScN layers.   

To summarize and conclude, we have used SE in a spectral range from 0.045 to 8.5\,eV to determine the optical constants of ScN layers grown on Al$_2$O$_3$ substrates using PAMBE. Parametric fits of the SE data return the energies of the four lowest direct band-to-band transitions of 2.03, 3.89, 5.33, and 6.95\,eV, corresponding to the high-symmetry critical points in the band structure of ScN. In the infrared range, free carrier absorption due to the high electron density dominates the dielectric function, and we have thus used Raman spectroscopy to examine the vibrational properties of the layers. Strong first-order modes are observed, which indicate the existence of defects in the layers breaking the symmetry of the rocksalt structure. Likewise, the shape and position of the PL band observed in the vicinity of the lowest direct gap is governed by defects relaxing $\mathbf{k}$ conservation and allowing transitions indirect in $\mathbf{k}$ space. The reduction of the electron density with increasing layer thickness suggests that O impurities stemming from the substrate are a likely candidate for these defects. A similar conclusion has been reached in Ref.~\citenum{Lupina2015Nov} for ScN growth on Sc$_2$O$_3$ buffer layers.\\

\small{See supplementary material for (1) symmetric $2\theta$--$\omega$ XRD scans of the ScN layers; (2) azimuthal scans performed in skew symmetry of the ScN\,220 and Al$_2$O$_3$\,11\=23 reflections; (3) evolution of the lattice constant with layer thickness, including literature values; (4) atomic force topographs of  layers with different thickness; (5) measured and fitted $\Delta$ and $\Psi$ spectra in the range of 0.045--0.15\,eV and the corresponding real and imaginary parts of the complex refractive index $\tilde{n}$ and the complex dielectric function $\epsilon$; (6) comparison of the real and imaginary parts of the complex refractive index $\tilde{n}$ and the complex dielectric function $\epsilon$ obtained for the 80- and 250-nm-thick ScN layers, compared to the results obtained by \citet{Maurya2022Jul}; (7) Drude and Tauc-Lorentz oscillators of the 80-nm-thick ScN layer; (8) peak energy and full-width at half-maximum of PL spectra of the layers as a function of electron density.\\
	
We thank Carsten Stemmler for expert technical assistance with the MBE system. 
}

\bibliography{BIB_Optics_ScN}

\end{document}


	
	\title{Optical properties of ScN layers grown on Al$_{2}$O$_{3}$(0001) by plasma-assisted molecular beam epitaxy}
	
	\author{Duc V. Dinh}
	\email[Electronic email: ]{dinh@pdi-berlin.de}
	\affiliation{Paul-Drude-Institut für Festkörperelektronik, Leibniz-Institut im Forschungsverbund Berlin e.V., Hausvogteiplatz 5--7, 10117 Berlin, Germany.}
	\author{Frank Peiris}
	\affiliation{Paul-Drude-Institut für Festkörperelektronik, Leibniz-Institut im Forschungsverbund Berlin e.V., Hausvogteiplatz 5--7, 10117 Berlin, Germany.}
	\affiliation{Department of Physics, Kenyon College, Gambier, 43022 Ohio, United States}
	\author{Jonas Lähnemann}
	\author{Oliver Brandt}
	\affiliation{Paul-Drude-Institut für Festkörperelektronik, Leibniz-Institut im Forschungsverbund Berlin e.V., Hausvogteiplatz 5--7, 10117 Berlin, Germany.}

\maketitle

\begin{figure}[t]
	\includegraphics[width=0.685\columnwidth]{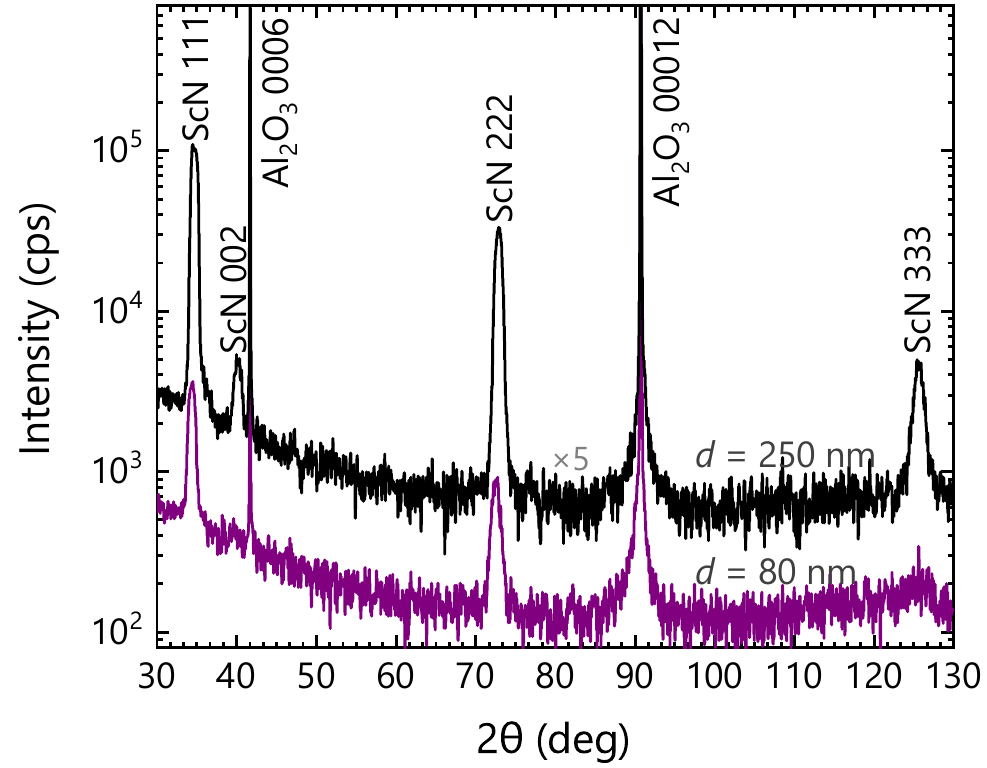}
	\caption{Wide-range symmetric $2\theta$--$\omega$ XRD scans of the ScN layers grown on Al$_2$O$_3$(0001) substrates with thicknesses of 80 and 250\,nm. Note the additional ScN\,002 reflection in the scan of the layer with $d$\,=\,250\,nm. These scans have been performed using a triple-axis high-resolution x-ray diffractometer (Philips Panalytical X'Pert PRO MRD) equipped with a two-bounce hybrid monochromator Ge(220) for the CuK$_{\alpha1}$ source ($\uplambda=\SI{1.540598}{\Angstrom}$).}
	\label{fig:XRD}
\end{figure}

\begin{figure}
	\centering
	\includegraphics[width=0.685\linewidth]{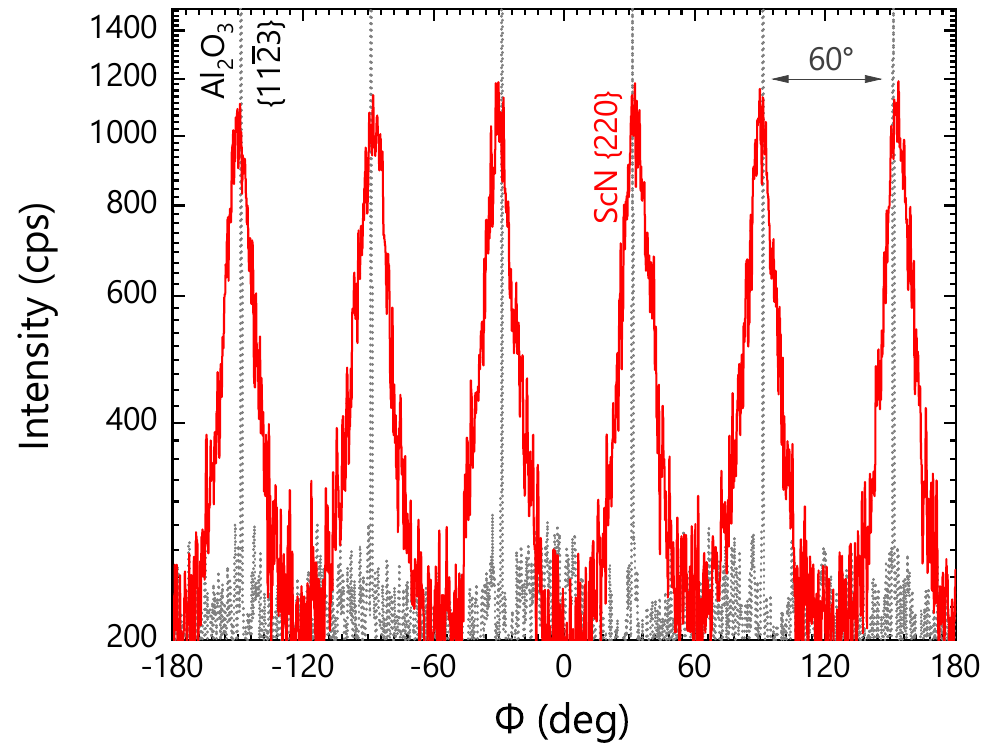}
	\caption{Azimuthal scans performed in skew symmetry of ScN\,220 ($\chi = 35.3$°) and Al$_2$O$_3$\,11\=23 reflections ($\chi = 61.2$°) reflecting the twinned growth of the ScN layers on Al$_2$O$_3$(0001).}
\end{figure}

\begin{figure}
	\centering
	\includegraphics[width=0.685\linewidth]{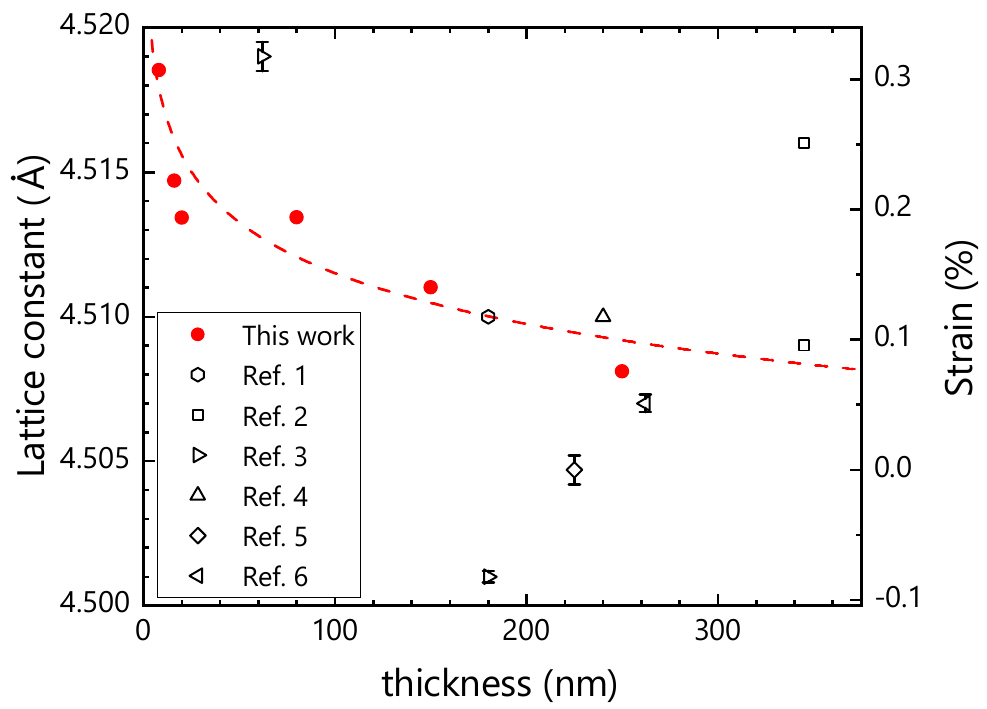}
	\caption{Dependence of the lattice constant of ScN layers on thickness. Values from the literature are shown for comparison (\citet{Gall1998Jul,Gall1998Dec,Gall1999Nov}, \citet{Al-Brithen2000Oct}, \citet{Moram2006Jul} and \citet{Deng2015Jan}). The strain values of the layers studied here are calculated by assuming that the relaxed lattice constant of ScN is of $\SI{4.5047}{\Angstrom}$, after Moram \textit{et al}.\cite{Moram2006Jul}}
\end{figure}

\begin{figure}[t]
	\includegraphics[width=\columnwidth]{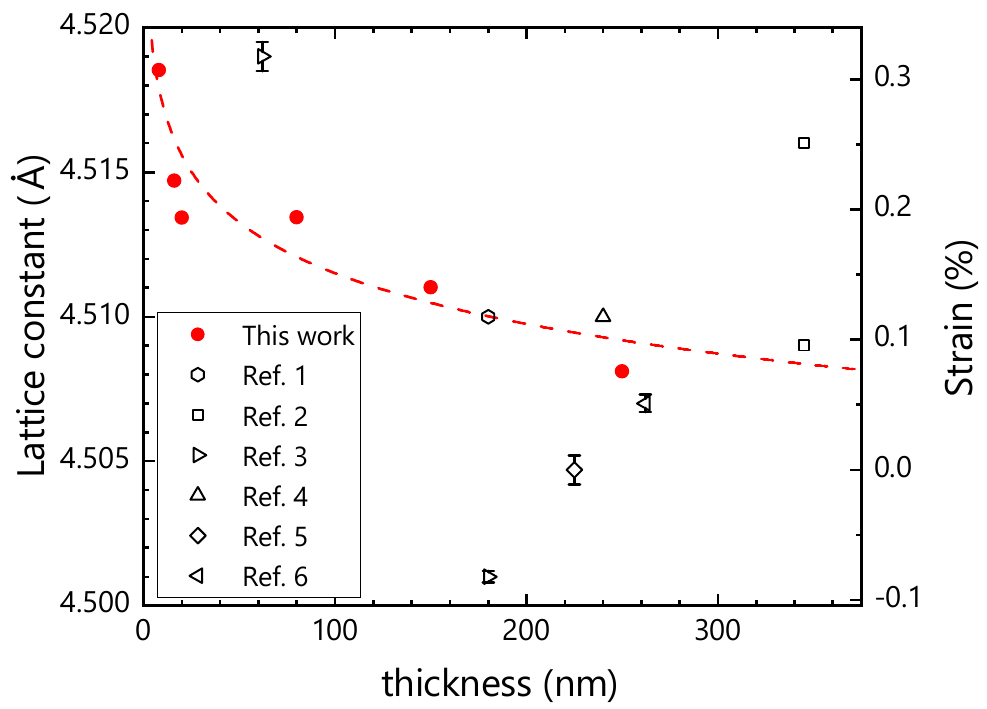}
	\caption{$2\,\times\,2$\,$\muup$m$^{2}$ atomic force topographs of ScN layers with thicknesses of 16, 80, and 250\,nm. The root-mean-square roughness determined from these topographs amounts to 0.2, 2.1, and 9.6\,nm, respectively. The topographs have been recorded by atomic force microscopy in contact mode (Dimension Edge, Bruker).}
	\label{fig:AFM}
\end{figure}

\begin{figure}[t]
	\includegraphics[width=\columnwidth]{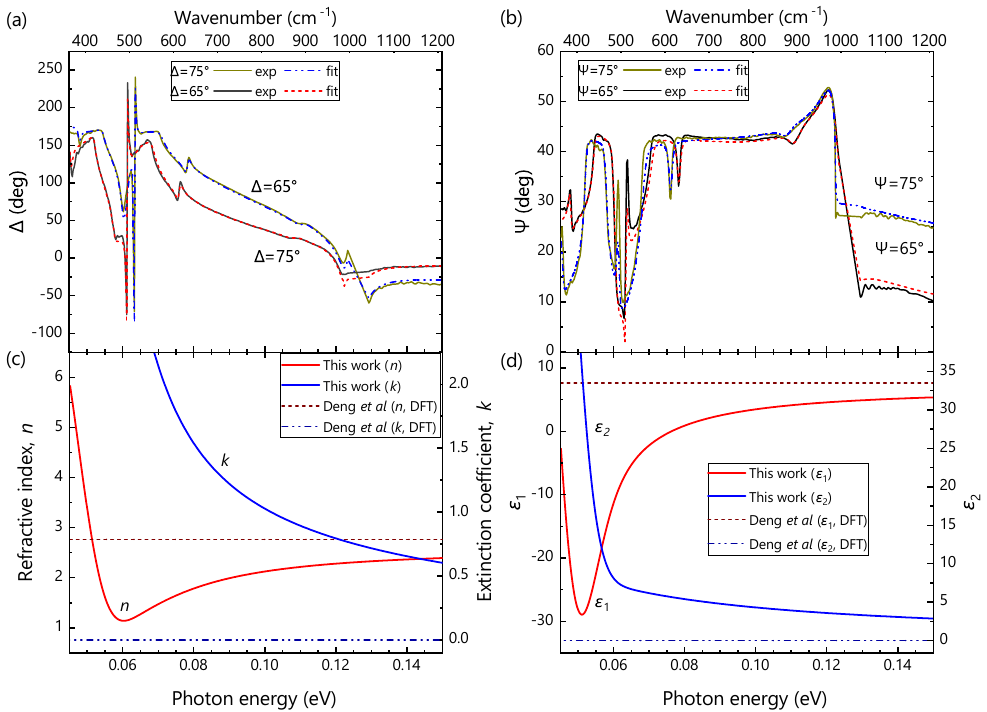}
	\caption{(a) $\Delta$ and (b) $\Psi$ spectra of the 80-nm-thick ScN layer and their point-to-point fits. (c) Real ($n$) and imaginary ($k$) parts of the complex refractive index as obtained from the fits. (d) Corresponding real ($\epsilon_1$) and imaginary ($\epsilon_2$) parts of the complex dielectric function. For comparison, (c) and (d) also show the theoretical values (DFT) obtained in Ref.~\citenum{Deng2015Jan}.}
	\label{fig:Ellip1}
\end{figure}

\begin{figure}[t]
	\includegraphics[width=\columnwidth]{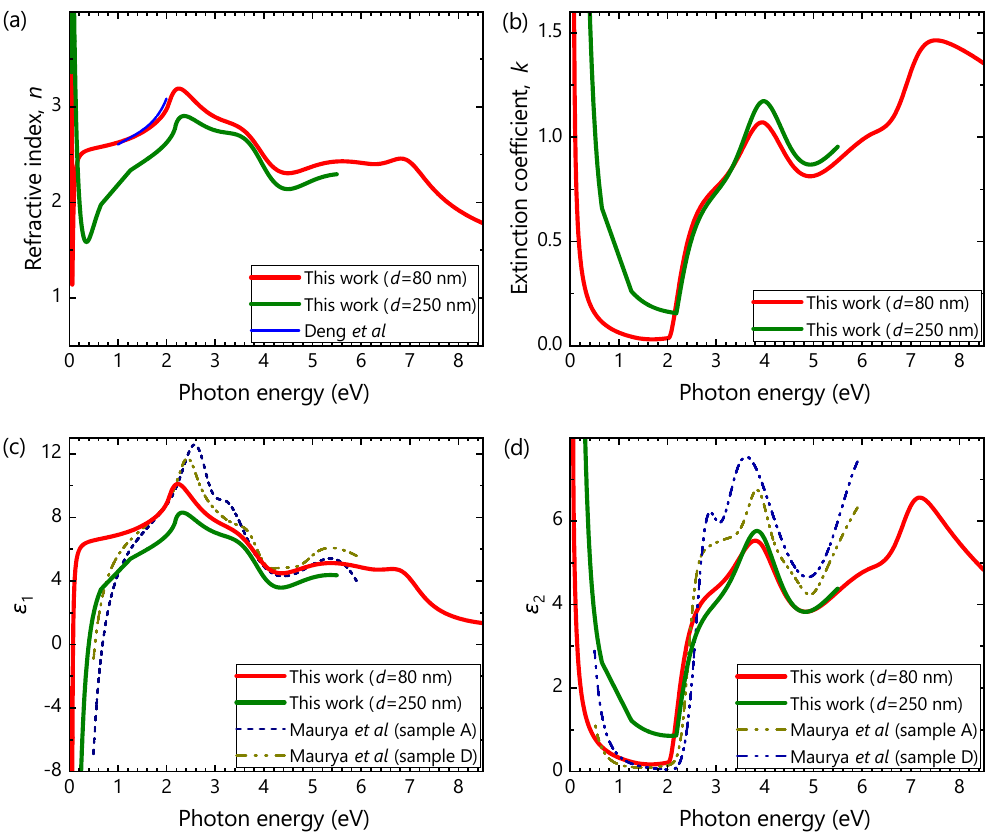}
	\caption{(a) Real ($n$) and imaginary ($k$) parts of the complex refractive index of the layers with $d$\,=\,80 and 250\,nm. Experimental curves taken from the work of \citet{Deng2015Jan} (sputtered ScN/MgO layers with thicknesses of 180--350\,nm) are plotted for comparison. Corresponding (c) real ($\epsilon_1$) and (d) imaginary ($\epsilon_2$) parts of the complex dielectric function. For comparison, spectra taken from the work of \citet{Maurya2022Jul} (sputtered ScN/TiN/MgO layers) are also included.}
	\label{fig:Ellip2}
\end{figure}

\begin{figure}[t]
	\includegraphics[width=0.685\columnwidth]{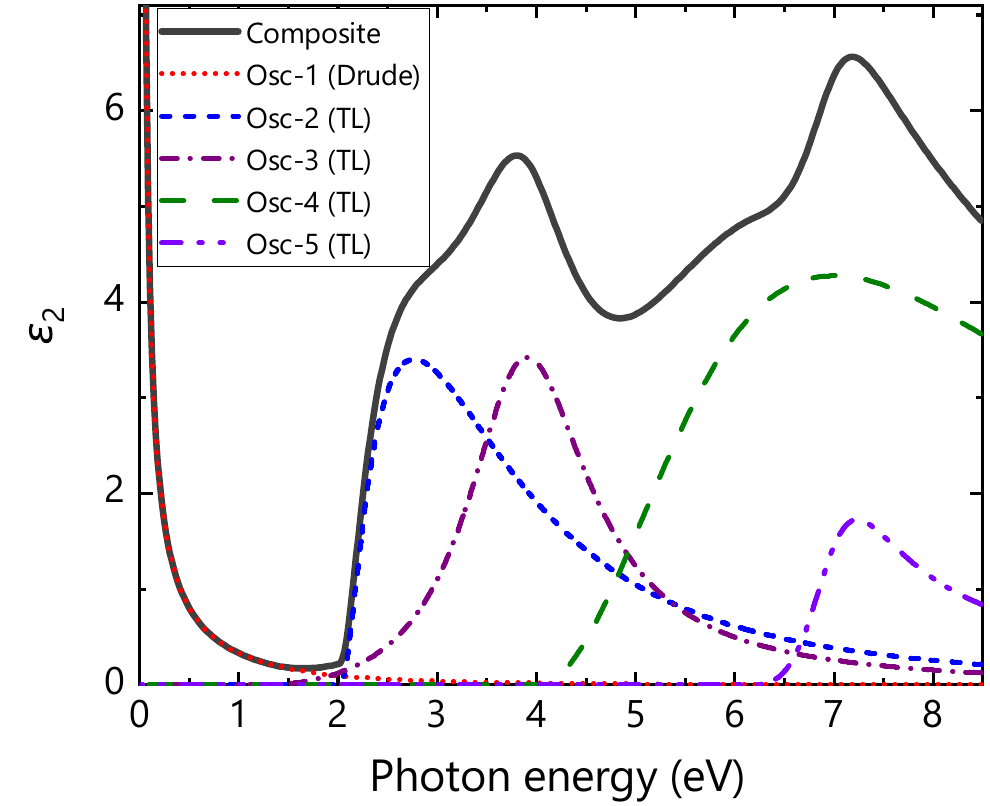}
	\caption{Drude and Tauc-Lorentz (TL) oscillators of the 80-nm-thick ScN layer.}
	\label{fig:Ellip3}
\end{figure}

\begin{figure}[t]
	\includegraphics[width=0.685\columnwidth]{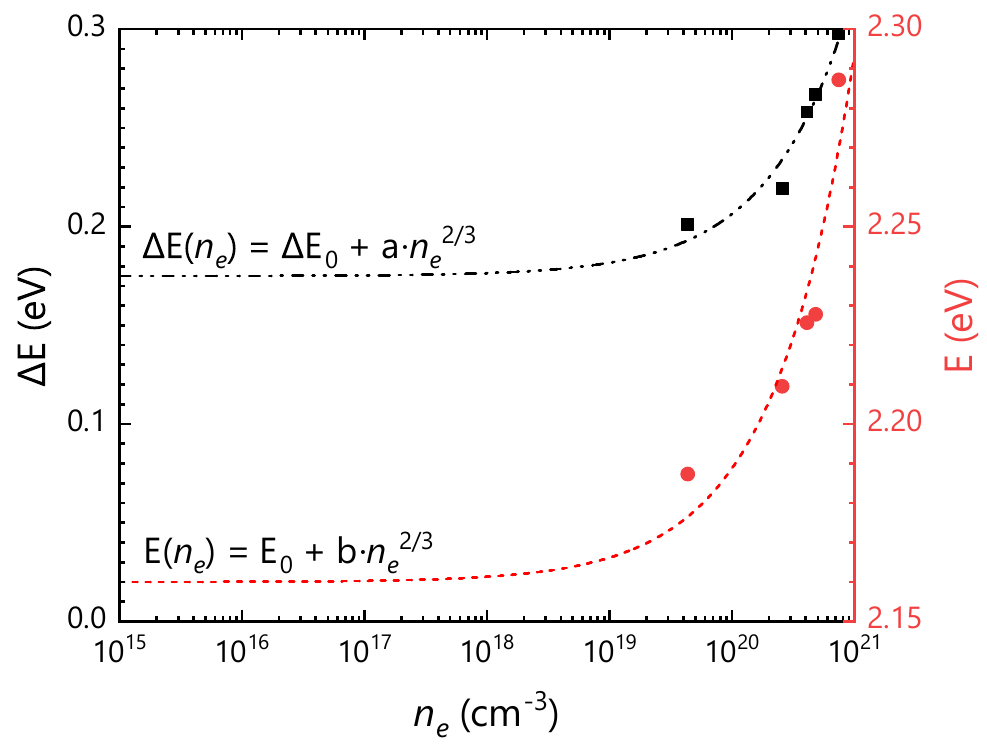}
	\caption{Full-width at half maximum ($\Delta E$) and peak energy ($E$) of room-temperature PL spectra of the ScN layers as a function of carrier concentration ($n_e$). The data are consistent with $n_e^{2/3}$ relation as indicated by the lines representing fits of the respective data sets as shown in this figure. This dependence indicates that both $E$ and $\Delta E$ are related to the Fermi level in the degenerately doped material. $E(n_e)$ would thus reflect the Burstein-Moss shift.\cite{Davydov2002Apr,Wu2002Nov,Lu2007Apr,Deng2015Jan,Mu2021Aug}}
	\label{fig:FWHM}
\end{figure}

	\clearpage
	\bibliography{BIB_Optics_ScN}